\documentclass[journal,10pt]{IEEEtran}
\usepackage{amsmath,amsfonts}
\usepackage{ amssymb }
\usepackage{algorithmic}
\usepackage{algorithm}
\usepackage{array}
\usepackage[caption=false,font=normalsize,labelfont=sf,textfont=sf]{subfig}
\usepackage{textcomp}
\usepackage{stfloats}
\usepackage{url}
\usepackage{verbatim}
\usepackage{graphicx}
\usepackage{dutchcal}
\usepackage{cuted} 
\usepackage{tikz, pgfplots}
\usepackage{pslatex}
\usepackage{float,multirow,multicol}
\usepackage{soul}

\usepackage[normalem]{ulem}
\usepackage{etoolbox}

\newcommand{\R}{\mathbb{R}}

\usetikzlibrary{calc}
\usetikzlibrary{spy}
\newlength{\spyimagewidth}
\newlength{\spyimageheight}
\DeclareMathOperator{\prox}{prox}
\DeclareMathOperator{\DB}{DB}
\DeclareMathOperator{\SSS}{S}
\newcommand{\lhs}{g}
\newcommand{\lms}{f}
\newcommand{\hms}{F}
\newcommand{\hhs}{G}
\DeclareMathOperator{\proxSR}{Prox_{SR}}
\DeclareMathOperator{\downSR}{Down}
\DeclareMathOperator{\downSSR}{SpecDown}
\DeclareMathOperator{\upSR}{BPUp}
\DeclareMathOperator{\upSSR}{SpecUp}
\DeclareMathOperator{\proxSSR}{Prox_{SSR}}
\DeclareMathOperator{\proxFus}{Prox_{Fus}}
\pgfplotsset{compat=1.18}

\begin{document}

\title{Model-Guided Network with Cluster-Based Operators for Spatio-Spectral Super-Resolution }

\author{Ivan Pereira-Sánchez, Julia Navarro, Ana Belén Petro, and Joan Duran
\thanks{The authors are with the Institute of Applied Computing and Community Code (IAC3) and with the Dept.~of Mathematics and Computer Science, Universitat de les Illes Balears, Cra.~de Valldemossa km.~7.5, E-07122 Palma, Spain (email: \{i.pereira, julia.navarro, anabelen.petro, joan.duran\}@uib.es).
}
\thanks{This study was funded by MCIN/AEI/10.13039/501100011033/ and by the European Union NextGenerationEU/PRTR via MaLiSat project TED2021-132644B-I00; and also by MCIN/AEI/10.13039/501100011033 and by “ERDF A way of making Europe”, European Union, via MoMaLiP project Grant PID2021-125711OB-I00. Also, we are grateful for the funding provided by the Conselleria de Fons Europeus, Universitat i Cultura (GOIB)  FPU2023-004-C. The authors gratefully acknowledge the computer resources at Artemisa, funded by the EU ERDF and Comunitat Valenciana and the technical support provided by IFIC (CSIC-UV).
}
}

\markboth{Journal of \LaTeX\ Class Files,~Vol.~X, No.~X, May~2025}%
{Shell \MakeLowercase{\textit{et al.}}: A Sample Article Using IEEEtran.cls for IEEE Journals}

\maketitle
\begin{abstract}
This paper addresses the problem of reconstructing a high-resolution hyperspectral image from a low-resolution multispectral observation. While spatial super-resolution and spectral super-resolution have been extensively studied, joint spatio-spectral super-resolution remains relatively explored. We propose an end-to-end model-driven framework that explicitly decomposes the joint spatio-spectral super-resolution problem into spatial super-resolution, spectral super-resolution and fusion tasks. Each sub-task is addressed by unfolding a variational-based approach, where the operators involved in the proximal gradient iterative scheme are replaced with tailored learnable modules. In particular, we design an upsampling operator for spatial super-resolution based on classical back-projection algorithms, adapted to handle arbitrary scaling factors. Spectral reconstruction is performed using learnable cluster-based upsampling and downsampling operators. For image fusion, we integrate low-frequency estimation and high-frequency injection modules to combine the spatial and spectral information from spatial super-resolution and spectral super-resolution outputs. Additionally, we introduce an efficient nonlocal post-processing step that leverages image self-similarity by combining a multi-head attention mechanism with residual connections. Extensive evaluations on several datasets and sampling factors demonstrate the effectiveness of our approach. The source code will be available at \url{https://github.com/TAMI-UIB/JSSUNet}. 
\end{abstract}

\begin{IEEEkeywords}
Joint spatio-spectral super-resolution, super-resolution, spectral super-resolution, image fusion, unfolding model, hyperspectral imaging. 
\end{IEEEkeywords}

\section{Introduction}

\begin{table}[t]
    \centering
    \caption{List of abbreviations and notations used throughout the paper
    }
    
    \begin{tabular}{c|c}
        \hline
        Symbol & Concept\\ 
        \hline
        LR &  Low-resolution \\
        HR &  High-resolution \\
        MS & Multispectral\\
        HS & Hyperspectral \\
        LR-MSI &  Low-resolution multispectral image\\
        HR-MSI &  High-resolution multispectral image\\
        LR-HSI &  Low-resolution hyperspectral image\\
        HR-HSI &  High-resolution hyperspectral image\\
        $\lms$ &  Observed LR-MSI\\
        $\hms$ &  Intermediate HR-MSI\\
        $\lhs$ &  Intermediate LR-HSI\\
        $\hhs$ & Reference HR-HSI\\
        $W$, $H$ & Width and height in high resolution\\
        $w$, $h$ & Width and height in low resolution\\
        $c$ & Number of multispectral bands\\
        $C$ & Number of hyperspectral bands\\
        $s$ & Sampling factor\\
        SR & Spatial super-resolution \\
        SSR & Spectral super-resolution \\
        SSSR & Joint spatio-spectral super-resolution \\
        Fus & Image Fusion \\
         \hline
    \end{tabular}
    \label{tab:notation}
\end{table}

Over the past decade, interest in hyperspectral (HS) imagery has grown due to its wide range of applications in fields such as remote sensing~\cite{wang2022remote}, agriculture~\cite{lu2020recent}, medical analysis~\cite{yoon2022hyperspectral}, and food safety~\cite{ravikanth2017extraction}, among others. Accurate estimation of the continuous radiation spectrum is essential for understanding the chemical-physical composition of objects, which is precisely the goal of HS imagery. Although current sensor devices are able to capture precise spectral signatures, they are often affected by noise due to the limited number of photons available within narrow bandwidths. Moreover, this constraint limits spatial resolution, as the sensor must sample over larger areas. Several approaches have been proposed to address this issue. For example, some satellite imaging systems~\cite{zhu2018review} simultaneously acquire a low-resolution hyperspectral image (LR-HSI) and a high-resolution multispectral image (HR-MSI), which are then fused to produce a high-resolution hyperspectral image (HR-HSI). However, these systems are typically expensive and present technical challenges, such as the need for precise band registration.

When only a single multispectral (MS) image is available, or in cases where the aforementioned limitations of multimodal data cannot be adequately addressed, joint spatio-spectral super-resolution (SSSR) approaches are required. The goal of SSSR is to reconstruct a HR-HSI from a low-resolution multispectral image (LR-MSI). Although this represents a cost-effective solution for many applications, it remains a highly ill-posed inverse problem.

Recent advances in deep learning have yielded promising results in HS imaging. Convolutional neural networks~\cite{li2017hyperspectral} have demonstrated the ability to capture complex spatial and spectral correlations in HS data. Attention mechanisms~\cite{hu2021hyperspectral} have been introduced to provide adaptability based on these correlations, as exemplified by the pioneering transformer layer~\cite{zhang2023essaformer}. Furthermore, model-driven methods~\cite{ma2021deep} have shown the importance of incorporating the physical properties of sensor devices when designing network architectures through unfolding strategies. While numerous advances have been made in HS image super-resolution, most existing methods primarily focus on enhancing either the spatial or the spectral resolution individually, or jointly improving both when two complementary observations are available.

The problem of SSSR was first addressed by Mei et al.~\cite{mei2020spatial}, who proposed a simple end-to-end network based on 3D convolutions. Despite this initial work, relatively few studies have continued to explore this direction. Ma et al.~\cite{ma2021deep} introduced a variational formulation in which the iterative solver is unfolded into a trainable network, replacing classical mathematical operations with learnable modules specifically designed for the task. In particular, they replace a matrix inversion computed in the low-resolution (LR) domain with a nonlocal layer. Subsequently, in \cite{ma2022multi}, the same authors proposed decomposing the SSSR problem into the three subproblems: spatial super-resolution (SR), spectral super-resolution (SSR), and fusion, each addressed with specific learning architectures. This decomposition has been shown to improve the final spatio-spectral super-resolved result. However, the use of networks without explicitly incorporating the physical modeling underlying each subproblem results in a less flexible and interpretable approach. In addition, while the nonlocal layer proposed by Ma et al.~leverages image self-similarities, its implementation incurs a high computational cost, and its application in the LR domain may result in suboptimal performance.

In this paper, we propose an end-to-end model-driven framework that explicitly decomposes the SSSR problem into SR, SSR, and image fusion. Each subtask is addressed using an interpretable deep learning architecture based on algorithm unfolding. Additionally, we introduce an efficient nonlocal post-processing module applied in the HR domain, where the attention map is restricted to only the top most similar pixels to balance accuracy and computational cost. Our main contributions are as follows:
\begin{itemize}
    \item In contrast to \cite{ma2021deep, ma2022multi}, and to the best of our knowledge, this is the first work that combines SR, SSR, and fusion using algorithm unfolding in an end-to-end manner.
    \item We adopt a single step of classical back-projection algorithms~\cite{irani1993motion,zomet2001robust} to construct a learnable architecture for the spatial upsampling operator, in which the design follows the prime factorization of the scaling factor.
    \item We introduce cluster-based spectral upsampling and downsampling operators that divide the image into clusters using a learnable module, and apply a multi-layer perceptron with tailored weights to compute the transformations for each cluster.
    \item In the fusion step, we adapt the radiometric constraint proposed in \cite{duran2017survey} to inject the geometry of the HR-MSI into the fused product. To this end, we introduce low-frequency estimation and high-frequency injection modules that operate on the images obtained from the SR and SSR stages.
    \item We propose an efficient post-processing module that leverages image self-similarities by combining multi-head attention mechanisms and residual networks. In addition, the attention weights are computed in a reduced embedding space and restricted to the top $10\%$ most similar pixels, boosting the performance while reducing the computational cost.
    \item We provide a comprehensive evaluation comparing all SSSR state-of-the-art methods with publicly available code, and demonstrate the superiority of our model across different datasets and sampling conditions. 
\end{itemize}

For clarity, the abbreviations of the most frequently used terms and the notations used throughout the paper are summarized in Table~\ref{tab:notation}. 

The rest of the paper is organized as follows. Section \ref{sec:sota} provides a comprehensive overview of related work in SR, SSR, image fusion, and SSSR. In Section \ref{sec:model}, we present the proposed variational formulation and the unfolded architecture for each subproblem, along with the involved operators and the post-processing module. Section \ref{sec:implementation} provides the implementation details of the training phase. In Section \ref{sec:experimentation}, we conduct quantitative and qualitative comparisons with existing techniques on several datasets under different sampling factors. Section \ref{sec:ablation} contains an ablation study assessing the impact of different components of the proposed method. Finally, conclusions and future work are outlined in Section \ref{sec:conclusions}.

\section{Related Work}\label{sec:sota}

\subsection{Spatial Super-Resolution}

Spatial super-resolution (SR) is a technique aimed at improving the pixel density and details of a LR image, producing a HR version that appears sharper and better defined. Over the years, researchers have introduced various SR techniques, which are commonly classified into four categories~\cite{lepcha2023image}: interpolation-based methods, model-based methods, learning-based methods, and transformer-based methods. 

Interpolation-based SR methods, such as bilinear and bicubic interpolation~\cite{li2001new, keys1981cubic} are fast and simple, making them a commonly used approach for image zooming due to their simplicity. These techniques work by estimating pixel intensities on an up-sampled image.  Although these methods are reliable and efficient for real-time SR applications, they often suffer from accuracy limitations.

Model-based methods~\cite{Khattab18ReconstructionReview} assume that the LR observed image is derived from the sought HR one after applying a sequence of operators, usually blur filtering and down-sampling. Reversing such a process is an ill-posed inverse problem, and thus prior knowledge on the structure of natural images must be assumed to regularize it. The most popular model-based methods are variational models, which define an energy functional that induces a high energy when the priors are not fulfilled~\cite{dong2011image, dong2012nonlocally, pereira2022if}. However, these methods need to define an explicit prior in the model-based formulation that comes from rigid assumptions about the nature of the underlying image.

To address the limitations of interpolation and model-based methods, learning-based SR techniques leverage data to infer prior information. Recent advances in deep learning have led to numerous successful deep learning-based SR approaches, particularly following Dong et al.~\cite{Dong2015ImageSuper}, who introduced a convolutional neural network (CNN) for SR. Since then, various mechanisms such as residual connections~\cite{Lim17EDSR}, dense connections~\cite{huang17DenseNet}, and generative adversarial networks (GAN)~\cite{Singla22GAN} have been used to improve performance. Recently, deep unfolding frameworks~\cite{dong2025dual} have been proposed to combine the strengths of both model-based and learning-based approaches, which unroll the steps of the optimization scheme derived from model-based formulations into a deep learning framework, resulting in more efficient and interpretable architectures.

We must give special consideration to the transformers-based methods \cite{dosovitskiy2020image,liang2021swinir,zhang2022efficient}. The attention mechanism allows transformers to capture long-range dependencies of the image. However, the quadratic cost associated with the number of projections and the huge amount of memory they require limits their capabilities for HR images.

Finally, various works specifically target the enhancement of the spatial resolution in the HS domain instead of the RGB domain. 
In this context,  Yuan et al.~\cite{Yuan17HSI} first introduced CNNs into the HS SR task. They formulated the SR goal as a transfer learning problem, where the mapping between low- and HR images is transferred from the RGB image domain to the HSI domain.  In \cite{jiang2020learning}, Jiang et al. propose SSPSR, which extracts several image features from overlapping groups of channels and subsequently fuses them to obtain the HR-HSI image. 

\subsection{Spectral Super-Resolution}

Spectral super-resolution (SSR) consists of enhancing the spectral resolution of images, particularly in the context of HS imaging. While traditional SR focuses on improving the spatial resolution of an image, SSR specifically targets its spectral bands. 
The main works in the literature focus on deep learning-based SSR approaches. The decisions on network architectures, feature extraction and the physical modeling are what fundamentally differentiate existing methods \cite{HE2023ReviewSSR}.

The architecture of the deep network plays a crucial role in learning-based methods. A commonly used architecture is UNet. This network consists of an 
encoder-decoder structure with skip connections to enhance both spatial and spectral details by extracting local and global features. Several works~\cite{Alvarez_Gila_2017,galliani2017learnedspectralsuperresolution,yan2018accuratespectralsuperresolutionsingle} have explored this architecture to tackle SSR. However, UNet's main drawbacks include difficulties in capturing extremely fine details or handling noisy data. 

The extraction of spatio-spectral features is central to SSR, as fully utilizing both spatial and spectral information improves the recovery of HS images. Effective approaches for spatio-spectral feature extraction include multi-feature fusion, 2D–3D methods, and attention mechanisms. The AWAN method \cite{li2020adaptive}, a representative model of attention mechanisms, achieves SSR by extracting shallow and deep features through a sequence of convolutional networks and residual blocks with attention layers.  Following this sequence, a patch-level second-order nonlocal module is applied, using the covariance matrix as attention maps, to capture long-range dependencies. 

Lastly, incorporating physical imaging models into algorithms is emerging as a key trend in SSR. Physical imaging models are mathematical or computational models that describe how the observed image is generated by the physical processes involved in imaging, such as sensor characteristics, camera optics, and image degradation. These models can help guide SSR algorithms by incorporating prior knowledge about the imaging system, which aids in recovering the true HS image from MS observations. Many studies have integrated physical modeling into CNNs, with different strategies like degradation simulation~\cite{Fubara20Degradation}, group recovery~\cite{Mei22group}, and model-embedded learning~\cite{Zhu1Embedded}.

\subsection{Image Fusion}

The goal of image fusion is to combine the spatial and spectral information from two data sources with complementary attributes. In this article, we consider the fusion of MS-HRI and HS-LRI, but the most common form of image fusion is pansharpening, which aims to fuse a gray-scale high resolution image called panchromatic (PAN) with a LR-MSI or LR-HSI. The applications of this field are numerous, although mostly related to remote sensing, since this problem appears with satellite images. Existing methods that tackle this problem~\cite{vivone2020new,VIVONE2023405,pereira2024comprehensive} can be roughly divided into classic and deep learning categories. 

Classic methods can be subclassified into Component Substitution (CS), Multi-Resolution Analysis (MRA), and Variational Optimization (VO). Component substitution methods aim to replace a specific component of the MS/HS image with the corresponding component extracted from the PAN image. For this purpose, several approaches have been proposed, such as Brovey \cite{gillespie1987color}, Principal Component Analysis (PCA) \cite{chavez1991comparison}, Intensity Hue Saturation (IHS) \cite{chavez1991comparison}, and Gram-Schmidt (GS) \cite{aiazzi2007improving}. MRA methods inject spatial details, acquired through multiscale decomposition of the PAN image, into the MS/HS data. The multiscale decomposition can be carried out by different techniques, such as the Laplacian pyramid \cite{aiazzi2006mtf}, the discrete wavelet transform \cite{king2001wavelet, otazu2005introduction}, or the high-pass modulation \cite{de1998fusion,khan2009pansharpening}.  Variational optimization methods address the fusion problem by minimizing an energy functional derived from observation models and prior knowledge of the expected solution. In this setting, several works can be found for pansharpening and hypersharpening based on different regularization and data terms \cite{ballester2006variational,duran2014nonlocal,duran2017survey,liu2024pansharpening}.

Deep learning methods can also be subclassified into pure learning methods and deep unfolding methods. In the case of pure learning-based works, several approaches have adopted residual architectures~\cite{Yang_2017_ICCV,he2019pansharpening,cai2020super} and attention mechanisms~\cite{lu2023awfln}, especially with transformer layers~\cite{bandara2022hypertransformer,zhou2022panformer}. For example, Fusformer~\cite{hu2022fusformer} achieves the fusion of images by employing a transformer-based encoder and decoder to capture long-term dependencies and extract geometric information. The encoder consists of one transformer layer, while the decoder consists of two. Each pixel is embedded using a linear layer and treated as a patch representation due to the richer spectral signature. Similarly to SR and SSR, the unfolding paradigm has been also explored for image fusion, based on variational formulations~\cite{mai2024deep} or on back-projection~\cite{zhang2023spatial} approaches. 

\subsection{Joint Spatio-Spectral Super-Resolution}

While SR, SSR and image fusion have been widely explored, only a few studies have tackled the problem of SSSR. The first to address this problem were Mei et al.~\cite{mei2020spatial}, who proposed a 3D convolution sequence within an end-to-end network. Subsequently, two works by Ma et al. further advanced the field, US3RN~\cite{ma2021deep} and SSFIN~\cite{ma2022multi}.

US3RN~\cite{ma2021deep} is based on a variational formulation specific to SSSR. The iterative minimization scheme determined by ADMM~\cite{boyd2011distributed} is unfolded by replacing the traditional operators with learnable components. The architecture combines a CNN with nonlocal modules to reconstruct the HR-HSI. In contrast, SSFIN~\cite{ma2022multi} proposes a network composed of three branches: one for SR, one for SSR, and one for image fusion. It extracts spatial and spectral features using residual layers with attention structures in the SR and SSR branches. The fusion branch then combines these features through residual blocks with dual attention components. 

Recently, Zhang et al.~\cite{zhang2024hyperspectral} introduced Implicit Neural Representation (INR) via the proposed LISSF, which consists of an encoder-decoder architecture. The encoder transforms the input into a 3D deep feature space, and the decoder maps each continuous spatial-spectral coordinate to the hyperspectral pixel value using a multi-layer perceptron. Finally, all HS pixel values are synthesized and reshaped to generate the resulting HR-HSI. 

\section{Proposed Model}\label{sec:model}

\begin{figure}[t]
    \centering
    \includegraphics[width=0.85\linewidth]{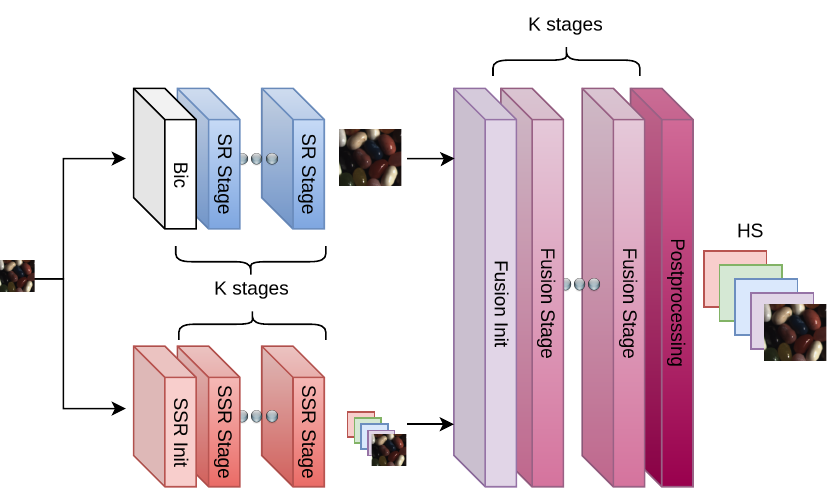}
    \caption{Overview of the proposed joint spectral and spatial super-resolution approach. The low-resolution multispectral image is first processed through two separate modules: a spatial super-resolution unit to produce a high-resolution multispectral image, and a spectral super-resolution unit to generate a low-resolution hyperspectral image. Finally, an image fusion module combines both estimations to obtain the desired high-resolution hyperspectral image. At each stage, one iteration of  \eqref{eq:gen_min_scheeme} is performed, using fidelity term and proximity operator specific to each task.}
    \label{fig:sssr_flow}
\end{figure}

Let $\lms\in \R^{h\times w \times c}$ denote the LR-MSI and $\hhs \in \R^{H\times W \times C}$ the HR-HSI to be estimated. We define $s=\frac{H}{h}=\frac{W}{w}$ as the sampling factor between the HR and LR domains. Additionally,  $c \ll C$ denotes the number of MS and HS channels, respectively. The goal of SSSR is to recover $\hhs$ from the observed image $\lms$. To address this problem, we decompose it into three tasks: SR, which provides a HR-MSI from $\lms$; SSR, which yields a LR-HSI from $\lms$; and the fusion of the resulting HR-MSI and LR-HSI to obtain the final product $\hhs$. Figure~\ref{fig:sssr_flow} illustrates the flow of the proposed approach.

 We tackle each subproblem by unfolding a variational formulation that combines a fidelity and a regularization term. In this setting, each minimization problem can be expressed as
 \begin{equation}
     \min_{u_i}  \mathcal{D}_i(u_i)  +  \lambda_i \mathcal{R}_{i}(u_i), \hspace{1cm}  i \in \mathcal{G},
 \end{equation}
where $\mathcal{G}=\{SR, SSR, Fus\}$ denotes the set of distinct goals. Here, $\mathcal{D}_i$ is the fidelity term, typically derived from an image observation model, $\mathcal{R}_i$ represents the regularization term, aimed at capturing the intrinsic properties of the image, and $\lambda_i$ is a trade-off parameter.  The minimum of this energy is obtained using the proximal gradient algorithm~\cite{parikh2014proximal}. The steps of the iterative scheme are
 \begin{equation}\label{eq:gen_min_scheeme}
     u_i^{n+1} = \prox_{\tau_i\lambda_i \mathcal{R}_{i}}(u_i^n -\tau_i \nabla \mathcal{D}_{i}(u_i^n)),
 \end{equation}
where $\prox$ denotes the proximity operator. 

We unfold \eqref{eq:gen_min_scheeme} by replacing the involved operators by learnable neural network architectures.  Regarding the proximity operator, if we assume that $\tau_i$ is sufficiently small and that $\lambda_i\mathcal{R}_i$ is differentiable, it can be approximated as follows~\cite{chambolle2016introduction}:
\begin{equation*}
    \prox_{\tau_i \lambda_i \mathcal{R_i}}(x_i) \approx x_i - \tau _i\lambda _i\nabla \mathcal{R}_i(x_i).
\end{equation*}
Accordingly, proximity operators are typically close to the identity, and thus residual architectures are a natural choice to replace them. Replacing the proximity operator with a neural network eliminates the need for an explicit regularization term. The intrinsic properties of the output image are implicitly learned from data during training.

Classical optimization schemes such as \eqref{eq:gen_min_scheeme} typically involve hundreds of iterations. However, due to the computational demands of training, unfolded methods commonly rely on a reduced number. To avoid confusion with training epochs, these optimization steps will be referred to as {\it stages}. All presented unfolding models are iterated over $K$ stages.

\begin{figure}[t]
    \centering
    \includegraphics[width=0.8\linewidth]{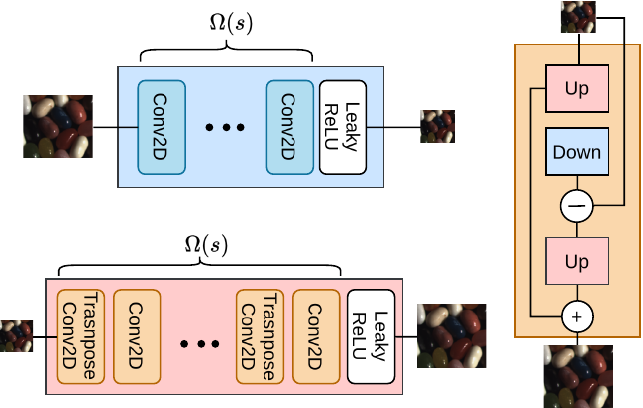}
    \caption
    {Architectures of the downsampling, $\downSR$, and upsampling, $\upSR$, operators involved in the spatial super-resolution task. $\downSR$ is shown in the top-left and consists of a sequence of 2D convolutions determined by the primer factor decomposition of the sampling factor, $\Omega(s)$. $\upSR$, displayed on the right, emulates one iteration of the back-projection algorithm~\eqref{eq:back-projection} by treating $\lms$ as the observed image and  $u^k$ as the upsampled input. Specifically, $\DB$ in \eqref{eq:back-projection} is replaced by $\downSR$, and $\kappa \ast \text{Bic}$ is replaced by $\text{Up}$, which is described in the bottom-left.}
    \label{fig:sr_up_and_down}
\end{figure}

\begin{figure}[t]
    \centering
    \includegraphics[width=0.8\linewidth]{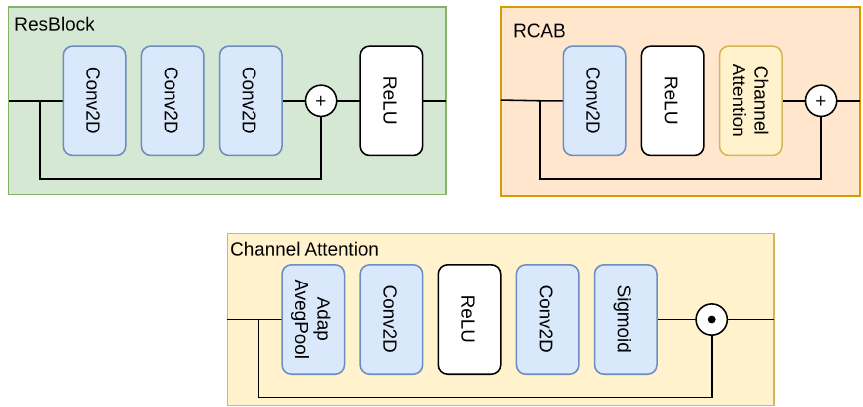}
    \caption{Architectures of two basic blocks used in all modules: the Residual Block (ResBlock) and the Residual Channel Attention Block (RCAB). The channel attention component used in RCAB is shown at the bottom, where the layer labeled {\it Adap AvegPool} refers to adaptive average pooling.}
    \label{fig:basic_blocks}
\end{figure}

In the following subsection, we discuss the modeling of each fidelity term and describe the architectures used to replace the involved operators. The index $i \in \{SR, SSR, Fus\}$ will be omitted, with the understanding that all parameters, variables, and operators are defined within the context of the respective subproblem.

\subsection{Spatial super-resolution subproblem}

This module aims to increase the spatial resolution of the LR-MSI while preserving the spectral information. To this end, we base our formulation on the classical observation model
\begin{equation}\label{eq:DBu-f}
    \lms= \DB(u) +\eta,
\end{equation}
where $\eta$ represents the noise realization and the operator $\DB$ is assumed to be a low-pass filter followed by an $s$-decimation operator. The corresponding fidelity term in \eqref{eq:DBu-f} is
\begin{equation*}
   \mathcal{D}(u) = \frac{1}{2}\| \DB (u)-f\|^2,
\end{equation*}
with its gradient given by
\begin{equation*}
   \nabla \mathcal{D}(u) = (\DB)^T (\DB (u)-\lms),
\end{equation*}
where $(\DB)^T$ denotes the adjoint operator of $\DB$. In this scenario, we substitute $\DB,  (\DB)^T$ and $\prox_{\tau\lambda \mathcal{R}}$ by the learnable modules $\downSR$, $\upSR$, and $\proxSR$, respectively. 

\begin{figure}[t]
    \centering
    \includegraphics[width=0.8\linewidth]{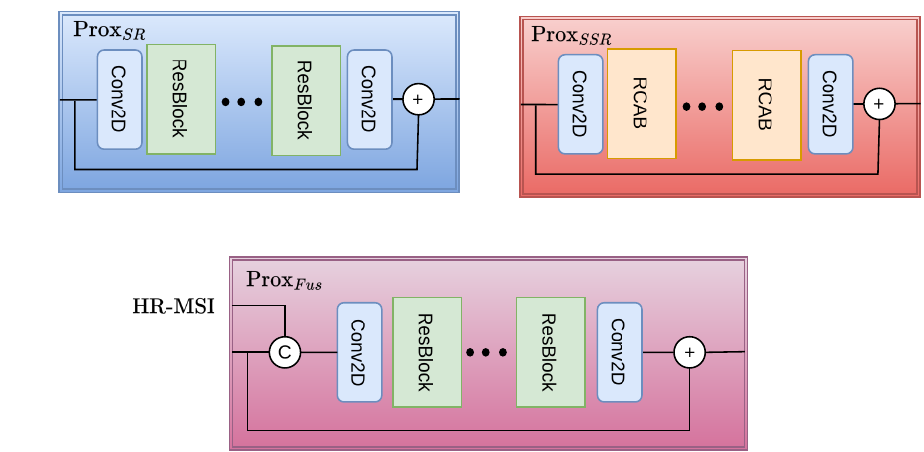}
    \caption{Architectures of the proximity operators $\proxSR$, $\proxSSR$, and $\proxFus$, respectively tailored to the task of improving spatial resolution, enhancing spectral resolution, and fusing images with complementary attributes. The ResBlock and RCAB modules are described in Figure \ref{fig:basic_blocks}.}
    \label{fig:proximity}
\end{figure}

The architecture of $\downSR$ consists of a sequence of 2D convolutions, as shown in Figure \ref{fig:sr_up_and_down}. This sequence is determined by the prime decomposition of the sampling factor, and each convolution uses a stride equal to the corresponding prime factor $p$ and a kernel size of $2p+1$. 

To mimic the classical formulation, the architecture of the upsampling module $\upSR$ should behave as the adjoint operator of $\DB$. However, empirical experiments (refer to Section~\ref{sec:ablation-sr} for more details) show that improved performance is achieved when using back-projection algorithms~\cite{irani1993motion,zomet2001robust}, a classical approach to SR that aims to minimize the reconstruction error between the target image and the output. In this process, the reconstructed image $u$ is obtained by iterating 
\begin{equation}\label{eq:back-projection}
u^{k+1} = u^k + \kappa \ast \text{Bic}(\DB(u^k)-\lms),
\end{equation}
where $\text{Bic}$ denotes the bicubic interpolation operator and $\kappa$ is a back-projection kernel that regulates the convergence speed. The proposed module $\upSR$, illustrated in Figure \ref{fig:sr_up_and_down}, emulates one iteration of this approach by treating $\lms$ as the LR input and  $u^k$ as the upsampled input obtained via the $\text{Up}$ operator, also depicted in Figure~\ref{fig:sr_up_and_down}. Specifically, $\DB$ is replaced by $\downSR$ and $\kappa \ast \text{BIC}$ is replaced by $\text{Up}$.

Finally, the learnable proximity operator $\proxSR$ consists of a sequence combining 2D convolutions and residual blocks, followed by a residual connection to the input of the module. The residual block is depicted in Figure~\ref{fig:basic_blocks} and the overall architecture of $\proxSR$ is shown in Figure~\ref{fig:proximity}. We initialize $u^0$ in \eqref{eq:gen_min_scheeme} by standard bicubic interpolation.

 \subsection{Spectral super-resolution subproblem}

\begin{figure}[t]
    \centering
    \includegraphics[width=0.8\linewidth]{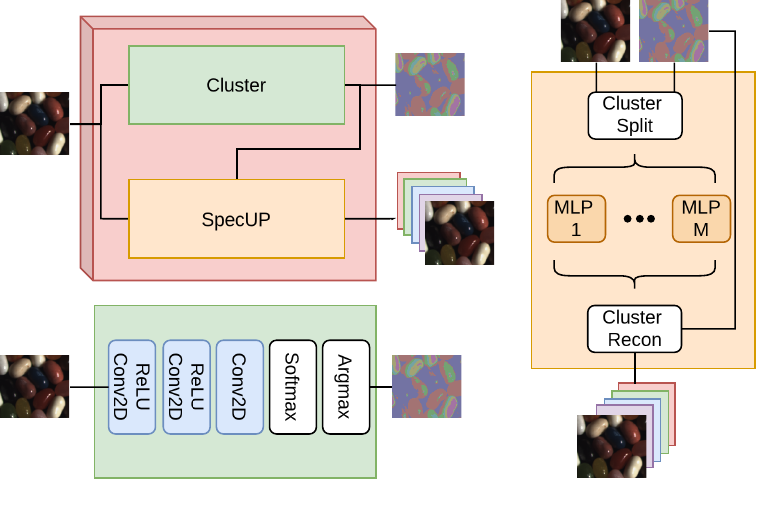}
    \caption{The diagram on the right illustrates the architecture of the spectral upsampling operator, $\upSSR$, where a tailored multilayer perceptron (MLP) is applied independently to each cluster. {\it Cluster Split} refers to the operation that separates and flattens the pixels within each cluster, while {\it Cluster Recon} denotes the rearrangement of the resulting pixels back into the image domain. The spectral downsampling operator, $\downSSR$, follows an analogous architecture, differing only in the MLP dimensions. The initialization step, which generates the clusters and the initial estimate $u^0$, is depicted in the top-left. The bottom-left shows the clustering module, where each pixel is assigned to the cluster with the highest associated probability.}
    \label{fig:ssr_init}
\end{figure}
 
 The goal now is to enhance the spectral resolution of the image while preserving the spatial information. The classical observation model is
 \begin{equation*}
    \lms  =\SSS(u) +\eta,
 \end{equation*}
where $\eta$ denotes the noise realization and the operator $\SSS$ represents the spectral downsampling, which is typically modeled as a linear combination of the spectral bands. Accordingly, the fidelity term is given by 
\begin{equation*} 
\mathcal{D}(u) = \frac{1}{2} \| \SSS(u) - \lms \|^2, 
\end{equation*}
and its gradient writes as
\begin{equation*} 
\nabla \mathcal{D}(u) = \SSS^T(\SSS(u) - \lms),
\end{equation*} 
where $\SSS^T$ denotes the adjoint operator of $\SSS$. In the unfolded framework, $\SSS,  \SSS^T$ and $\prox_{\tau\lambda \mathcal{R}}$ are replaced by the learnable modules $\downSSR$, $\upSSR$, and $\proxSSR$, respectively. 

Defining $\downSSR$ and $\upSSR$ is challenging since the linearity does not strictly hold in practice. To address this, we assume that $\downSSR$ and $\upSSR$ should vary for pixels with different spectral responses but remain consistent for those with similar spectral characteristics. Although this assumption does not ensure linearity, it facilitates approximating the operators within spectral response ranges. 

Accordingly, we divide the image into $M$ clusters (in practice, $M=10$), computed from $\lms$ using a learnable module. This module first applies a sequence of 2D convolutions, producing for each pixel a representation in an $M$-dimensional space (the cluster space). A $\text{Softmax}$ function is then applied across the cluster dimension to yield the probability of each pixel belonging to each cluster. Finally, an $\text{argmax}$ operation assigns each pixel to the cluster with the highest probability. The clusters are computed  during the initialization step and are used to construct $\downSSR$ and $\upSSR$.

Both spectral operators, $\downSSR$ and $\upSSR$, begin by splitting and flattening the pixels according to their assigned clusters. The pixels in each cluster are  then transformed into either the MS or HS space using a Multi-Layer Perceptron (MLP) with cluster-specific weights. The transformed pixels are then rearranged back into the image domain. Figure~\ref{fig:ssr_init} illustrates the architectures of these spectral downsampling and upsampling operators, as well as the initialization module.
 
 Finally, the architecture of $\proxSSR$ employs the Residual Channel Attention Block (RCAB) \cite{zhang2018image}, which incorporates a channel attention mechanism to enhance spectral information. A diagram of RCAB is shown in Figure \ref{fig:basic_blocks}, while the overall architecture of $\proxSSR$ is depicted in Figure \ref{fig:proximity}. We initialize $u^0$ in \eqref{eq:gen_min_scheeme} by $u^0 = \upSSR(\lms)$.
 
\subsection{Fusion subproblem}

Image fusion aims to generate a single HR-HSI by accurately combining the spatial details from the HR-MSI and the spectral content from the LR-HSI. These inputs, denoted by $u_{SR}$ and $u_{SSR}$, are produced by the preceding spatial and spectral super-resolution modules, respectively.

A particular case of fusion that has been extensively studied in remote sensing is pansharpening (or hypersharpening), which fuses a HR grayscale image with a LR-MSI (or a LR-HSI). Several fidelity terms have been proposed in the literature~\cite{pereira2024comprehensive}. In this work, we adapt the constraint introduced in \cite{duran2017survey} to inject the geometry from the HR-MSI into the fused product, while preserving the chromaticity of the LR-HSI. 

Let $\overline{u}_{SR}$ and $\overline{u}_{SSR}$ represent the low frequencies of the HR-MSI and the LR-HSI, respectively, and let $\hat{u}_{SR}$ be the high frequencies of the HR-MSI. The proposed constraint is formulated as
\begin{equation}\label{eq-const}
u-\overline{u}_{SSR}= (\overline{u}_{SSR}\oslash \overline{u}_{SR}) \odot (\hat{u}_{SR}-\overline{u}_{SR}),
\end{equation}
where $\odot$ and $\oslash$ denote the element-wise product and quotient, respectively. Therefore, $u-\overline{u}_{SSR}$ captures the high-frequency details of the fused product, $\hat{u}_{SR}-\overline{u}_{SR}$ contains those from the HR-MSI, which is assumed to encode the geometry, and $\overline{u}_{SSR}\oslash \overline{u}_{SR}$ acts as a modulation coefficient that compensates for differences in energy levels between MS and HS data. By multiplying both sides of equation \eqref{eq-const} by $\overline{u}_{SR}$ and rearranging, we obtain the fidelity term
\begin{equation*} 
\mathcal{D}(u)=\frac{1}{2}\|u\odot\overline{u}_{SR} - \overline{u}_{SSR}\odot\hat{u}_{SR}\|^2,
\end{equation*}
the gradient of which is
\begin{equation*}
    \nabla \mathcal{D}(u) = \overline{u}_{SR} \odot (u\odot\overline{u}_{SR} - \overline{u}_{SSR}\odot \hat{u}_{SR}).
\end{equation*}

\begin{figure}[t]
    \centering
    \includegraphics[width=0.8\linewidth]{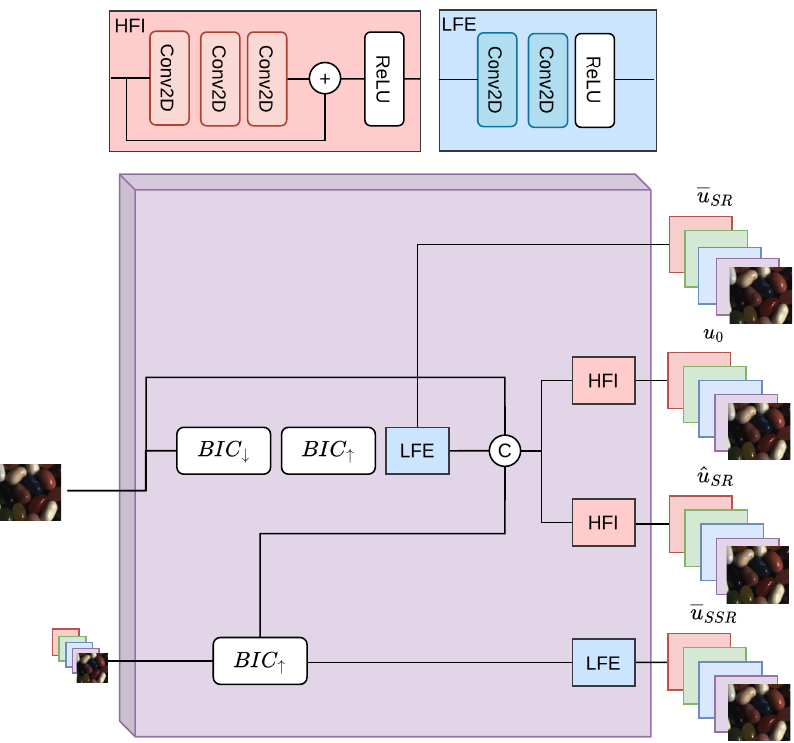}
    \caption{Architectures of the High Frequency Injection (HFI) module (top-left), the Low Frequency Estimation (LFE) module (top-right), and the initialization module (bottom) for the image fusion task. In this setting, $BIC_{\downarrow}$ refers to downsampling by bicubic interpolation, while $BIC_{\uparrow}$ refers to upsampling by bicubic interpolation. Parameters are not shared across different instances of the LFE and HFI modules.
    }
    \label{fig:fusion_init}
\end{figure}

The low-frequency information encoded in $\overline{u}_{SSR}$ is estimated using a sequence of convolutional layers applied to the bicubic interpolation of $u_{SSR}$. This process is referred to as the Low Frequency Estimation (LFE) module. Similarly, $\overline{u}_{SR}$ is estimated by applying the same module to a low-pass version of $u_{SR}$, obtained through downsampling followed by bicubic interpolation. Conversely, the high-frequency components in $\hat{u}_{SR}$ are extracted from $u_{SR}$ using residual blocks, a process referred to as the High Frequency Injection (HFI) module.  This module receives as input the concatenation of $u_{SR}$, the bicubic interpolation of $u_{SSR}$, and $\overline{u}_{SR}$. Both LFE and HFI modules produce their outputs in the HS domain to match the dimensions required for the pixel-wise product. The initial estimate $u^0$ is obtained using the same procedure as for $\hat{u}_{SR}$. The architectures of the LFE and HFI modules, along with the initialization step, are illustrated in Figure~\ref{fig:fusion_init}.

The proximity operator $\proxFus$ is designed to be guided by the geometry of the HR-MSI. We adopt an architecture analogous to that of $\proxSR$, with the difference that we concatenate the argument in \eqref{eq:gen_min_scheeme} with $u_{SR}$. The architecture of $\proxFus$ is shown in Figure~\ref{fig:proximity}.

\subsection{Post-processing}

\begin{figure}[t]
    \centering
    \includegraphics[width=0.8\linewidth]{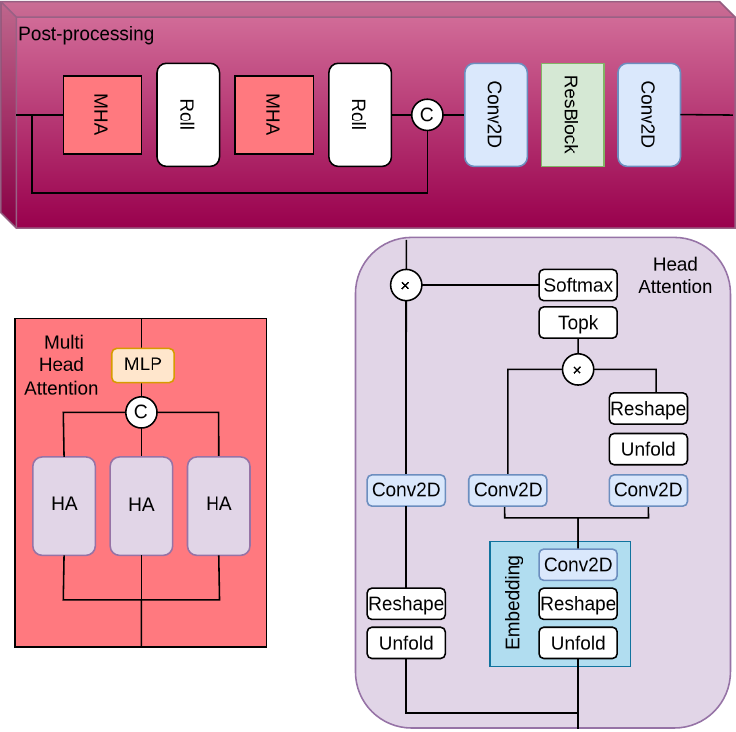}
    \caption{
    Proposed post-processing architecture (top), Multi-Head Attention mechanism (bottom-left) and Head Attention unit (bottom-right). In the Head Attention unit, the patch embedding $E$, highlighted by the blue square, is introduced to reduce the dimensionality for computing attention weights, and Topk refers to the strategy of selection only the top $10\%$ most similar pixels within each neighborhood. Both mechanisms are designed to boost computational efficiency while allowing the module to capture long-range dependencies. In the post-processing diagram, the MHA is applied twice: first in the original domain, and then to the image shifted by half the window size, a process denoted as Roll.
    }
    \label{fig:postprocessing}
\end{figure}

Finally, to correct potential defects and further enhance image geometry and chromaticity, we introduce a nonlocal residual architecture as a post-processing module. Specifically, we leverage image self-similarity using a Multi-Head Attention (MHA) mechanism, which applies a Head Attention unit multiple times with learnable parameters. Head Attention typically divides the image into non-overlapping patches and replaces each of the, with a weighted sum of all patches. In contrast, our proposed Head Attention operates at pixel level, computing a weighted sum over a neighborhood centered at each pixel. The extent of this neighborhood is referred to as the window size. While this restriction efficiently reduces the number of operations, it also limits the ability to capture long-range dependencies. To mitigate this limitation, we use larger window sizes ($11\times11$ in practice), but considering only the top 10\% most similar pixels within each neighborhood.

In order to compute pixel similarities effectively, we take into account a whole patch surrounding each pixel. Therefore, given a patch of size $P\times P$, with $P\in \mathbb{N}$ (we set $P=11$ in practice), we define the patch-image representation $X\in \mathbb{R}^{H \times W \times (C \cdot P^2)}$ based on the original image \(x \in \mathbb{R}^{H \times W \times C}\). Specifically, for each $ j \in \{1, \ldots, H\} \times \{1, \ldots, W\}$, we define $X_j = \bigl(x_l\bigr)_{l \in L_j}$, where $L_j = \bigl\{\,l \in \mathbb{Z}^2 : \|l - j\|_\infty \le \frac{P}{2} \bigr\}$.
For any index \(l \notin \{1,\ldots,H\} \times \{1,\ldots,W\}\), the corresponding value is set to zero.

Although incorporating pixel blocks provides valuable spatial context, it also introduces redundant information and increases computational cost. To address this issue, we introduce a patch embedding \(E(\cdot)\), which maps each patch to a low-dimensional representation via a 2D convolution. This results in the image representation $E(x) \in \mathbb{R}^{H\times W\times e_d}$, which is ultimately used to compute similarities between pixels. Here, \(e_d\) denotes the patch embedding dimension, which is set to $e_d=8$ in practice.

Finally, due to the use of zero padding when the neighborhood does not fully lie within the image domain, the MHA may introduce artifacts at the borders of the image. To alleviate this effect, we apply the filter twice. First, in the original domain and, second, after shifting the image by half the window size. The architecture of the proposed post-processing module is illustrated in Figure~\ref{fig:postprocessing}. 

\section{Implementation details}\label{sec:implementation}

The model was trained in two phases. In the first phase, the SR, SSR, and fusion modules were trained without applying the post-processing step. In the second phase, only the parameters of the post-processing were optimized, while the parameters of the other modules remained frozen. The training was conducted over 1000 epochs for the first phase and an additional 100 epochs for the second phase.

Each unfolding scheme was iterated over $K=4$ stages without sharing parameters between them, and the number of features of the residual blocks was set to 128 for each proximity operator. In Section~\ref{sec:ablation}, we report the effect of using different values for these parameters. 

The loss function used for training in the first phase includes one term for each stage and module, as well as an additional term for the final result. These terms are grouped by modules and weighted by hyperparameters. Thus, the loss function involves the reference HR-MSI $\hms$,  the LR-HSI $\lhs$, and  the HR-HSI $\hhs$:

\begin{equation}\label{eq:loss}
    \begin{split}
        \mathcal{L}_1(\{u^k_{i}\}_{k=1,i\in \mathcal{G}}^{K}, \hms, \lhs, \hhs) ) 
        =&\mathcal{l}_1(u_{Fus}^K, \lhs)+\frac{\alpha_{SR}}{K}\sum_{k=1}^K \mathcal{l}_1(u_{SR}^k, \hms), \\
        &+\frac{\alpha_{SSR}}{K}\sum_{k=1}^K \mathcal{l}_1(u_{SSR}^k, \lhs)\\
        &+\frac{\alpha_{Fus}}{K}\sum_{k=1}^K \mathcal{l}_1(u_{Fus}^k, \hhs)
    \end{split}
\end{equation}
where $u_{i}^k$ denotes the output at stage $k$ for the goal $i$, and $\alpha_{i}$ stands for the hyperparameter that weights the corresponding terms. 

During training, the loss hyperparameters were updated at 300 and 600 epochs. Specifically, the ordered values for initialization, 300 epochs, and 600 epochs are given by $\alpha_{SR}=\{2,0.5,0\}$, $\alpha_{SSR}=\{1,1,0.5\}$ and $\alpha_{Fus}=\{0.5,1,1\}$. These values were determined experimentally; see Section~\ref{sec:ablation} for more details. 

In the second phase, the loss function was the $l_1$ loss between the output and the reference image. In both phases, the ADAM optimizer~\cite{kingma2014adam} was used, with a learning rate of 0.0001.

\section{Experimentation}\label{sec:experimentation}

In this section, we evaluate the performance of the proposed SSSR method by analyzing both numerical and visual results. To provide a fair evaluation, we conduct experiments on various datasets, testing the generalization capabilities of the approaches.
Specifically, we use three different datasets, CAVE~\cite{yasuma2010generalized}, Pavia\footnote{Pavia dataset available at \url{https://www.ehu.eus/ccwintco/index.php/Hyperspectral_Remote_Sensing_Scenes#Pavia_Centre_and_University}}, and NTIRE2020~\cite{arad2020ntire}.

In the literature, there are few methods specifically designed for SSSR with publicly available source code. In particular, we only find SSFIN~\cite{ma2022multi} and US3RN~\cite{ma2021deep}, the results of which we include in our comparisons. To incorporate additional possibilities, we also compare with the combination of state-of-the-art techniques dealing with each individual objective. That is, we also evaluate the combination of the SR method SSPSR~\cite{jiang2020learning}, the SSR approach AWAN~\cite{li2020adaptive} and the image fusion method Fusformer~\cite{hu2022fusformer}. We test three different possibilities. First, applying SSPSR and then AWAN, which we denote as SR-SSR. Second, applying first AWAN and then SSPSR, denoted as SSR-SR. Third, we apply SSPSR and AWAN in parallel, and fuse the obtained super-resolved results using Fusformer, indicated as SR/SSR-Fus.

Quantitative evaluation has been conducted with the following reference metrics: ERGAS (Erreur Relative Globale Adimentionelle de Synth\`ese)~\cite{rabbani1991digital}, which measures the global spatial quality; PSNR (Peak Signal to Noise Ratio)~\cite{rabbani1991digital}, which assesses the spatial reconstruction quality; SSIM (Structural Similarity Index Measure)~\cite{wang2002universal}, which evaluates the overall quality of the fused image; and SAM (Spectral Angle Mapper)~\cite{pereira2024comprehensive},
which measures the spectral reconstruction quality. Note that for PSNR and SSIM the higher the better, while for ERGAS and SAM lower results are preferred. 

All compared techniques were trained using the loss functions and training hyperparameters specified in their respective articles. For all methods, the weights used for the evaluation correspond to the parameters that achieved the best performance in terms of PSNR on the validation set.

\begin{table*}[!ht]
\centering
\caption{Quantitative metrics on test images from the CAVE, NTIRE2020, and Pavia Center datasets. Methods were evaluated using sampling ratios of 2, 4, and 8. Bold indicates the best result; italic indicates the second best.} \label{tab:results}
\begin{tabular}{| l l | c c c c | c  c c c | c  c c c |}
\hline
 &  & \multicolumn{4}{c|}{\textbf{Sampling 2}} & \multicolumn{4}{c|}{\textbf{Sampling 4}} & \multicolumn{4}{c|}{\textbf{Sampling 8}} \\
 & \textbf{Model} & PSNR$\uparrow$ & SSIM$\uparrow$ & SAM$\downarrow$ & ERGAS$\downarrow$ & PSNR$\uparrow$ & SSIM$\uparrow$ & SAM$\downarrow$ & ERGAS$\downarrow$ & PSNR$\uparrow$ & SSIM$\uparrow$ & SAM$\downarrow$ & ERGAS$\downarrow$ \\
\hline
\multirow{6}{*}{\rotatebox[origin=c]{90}{CAVE}} & SR-SSR & 38.02 & 0.9648 & 10.65 & 7.20 & 34.65 & 0.9363& 15.57 & 10.10 & 32.92 & 0.8597 & 15.42 & 12.25 \\
 & SSR-SR & 38.22 & 0.9656 & 10.26 & 7.13 & 36.25 & 0.9376& 11.46 & 9.07 & 33.78 & 0.8568 & 13.64 & 12.20 \\
 & SR/SSR-Fus & 38.01 & 0.9648 & 10.36 & 7.31 & 35.91 & 0.9223& 12.28 & 9.10 & 33.73 & 0.8637 & 13.50 & 12.15 \\
 & SSFIN & \textit{38.92} & \textit{0.9709} & \textit{8.83} & 6.98 & \textit{37.02} & 0.9289& \textit{9.26} & 8.75 & 33.86 & 0.8651 & 13.53 & \textit{11.45} \\
 & US3RN & 38.57 & 0.9697 & 9.30 & \textit{6.86} & 36.86 & \textit{0.9278}& 9.34 & \textit{8.62} & \textit{33.97} & \textit{0.8764} & \textit{11.01} & 11.64 \\
 & Ours  & \textbf{39.39} & \textbf{0.9718} & \textbf{8.51} & \textbf{6.41} & \textbf{37.40} & \textbf{0.9427}& \textbf{8.96} & \textbf{8.28} & \textbf{34.60} & \textbf{0.8831} & \textbf{9.89} & \textbf{11.13} \\
\hline
\multirow{6}{*}{\rotatebox[origin=c]{90}{NTIRE2020}} & SR-SSR & 36.38 & 0.9578 & 3.52 & 2.27 & 31.48 & 0.8263 & 4.17 & 4.01 & 28.46 & 0.7182 & 4.87 & 5.71 \\
 & SSR-SR & \textit{38.40} & \textit{0.9803} & 3.03 & \textit{1.84} & \textit{33.80} & \textit{0.8886} & \textit{3.24} & \textbf{3.12} & \textit{29.76} & \textit{0.7516} & \textit{3.84} & \textit{5.06} \\
 & SR/SSR-Fus & 38.19 & 0.9796 & 3.14 & 1.86 & 33.37 & 0.8860 & 3.79 & 3.21 & 29.49 & 0.7494 & 4.46 & \textit{5.06} \\
 & SSFIN & 37.87 & 0.9740 & 3.34 & 2.14 & 32.66 & 0.8661 & 4.44 & 3.63 & 28.75 & 0.7295 & 5.44 & 5.85 \\
 & US3RN & 37.15 & 0.9617 & 3.23 & 2.49 & 31.40 & 0.8225 & 3.91 & 4.22 & 27.41 & 0.6932 & 6.02 & 6.78 \\
 & Ours  & \textbf{39.42} & \textbf{0.9819} & \textbf{2.42} & \textbf{1.74} & \textbf{34.13} & \textbf{0.8901} & \textbf{2.96} & \textit{3.15} & \textbf{29.80} & \textbf{0.7541} & \textbf{3.49} & \textbf{4.64} \\
\hline
\multirow{6}{*}{\rotatebox[origin=c]{90}{Pavia}}  & SR-SSR & 31.02 & 0.8783 & 6.26 & 5.04 & 27.64 & 0.7714& 7.28 & 7.20 & 25.03 & 0.5466 & 8.16 & 9.45 \\
 & SSR-SR & 31.71 & 0.9013 & 5.97 & 4.71 & 28.08 & 0.7968& 6.78 & 6.86 & 25.04 & 0.5544 & 8.16 & 9.41 \\
 & SR/SSR-Fus & 31.63 & 0.8911 & 5.29 & 4.61 & 27.73 & 0.7242& 6.89 & 6.98 & 25.05 & 0.5474 & 8.21 & \textit{9.25} \\
 & SSFIN & 34.09 & 0.9339 & 4.46 & 3.65 & 28.55 & 0.7536& 6.29 & 6.40 & \textit{25.12}& 0.5587 & 8.30 & 9.38 \\
 & US3RN & \textit{34.39} & \textit{0.9378} & \textit{4.35} & \textit{3.54} & \textit{29.09} & \textit{0.7278}& \textit{5.84} & \textit{6.17} & \textit{25.12} & \textit{0.5563} & \textit{8.09} & 9.29 \\
 & Ours  & \textbf{34.96} & \textbf{0.9451} & \textbf{3.98} & \textbf{3.31} & \textbf{29.46} & \textbf{0.8129}& \textbf{5.44} & \textbf{5.87} & \textbf{25.49} & \textbf{0.5864} & \textbf{7.44} & \textbf{8.96} \\
\hline
\end{tabular}

\end{table*}

\subsection{CAVE}


\newcommand{\spyimageCAVE}[2][0.22\linewidth]{%
  \setlength{\spyimagewidth}{#1}
  \setlength{\spyimageheight}{\spyimagewidth}
  \begin{tikzpicture}[spy using outlines={rectangle, magnification=2, size=0.4\spyimagewidth, connect spies,color=white}]
    \node[inner sep=0pt, anchor=south west] (image) at (0,0) {\includegraphics[width=\spyimagewidth]{#2}};
    \coordinate (spy point) at (0.5\spyimagewidth , 0.75\spyimageheight );
    \coordinate (spy node) at (0.2\spyimagewidth, 0.2\spyimageheight);
    \spy on (spy point) in node at (spy node);
  \end{tikzpicture}%
}

\begin{figure*}[t]
    \centering
\begin{tabular}{c@{\hskip 0.2em} c@{\hskip 0.2em} c@{\hskip 0.2em} c@{\hskip 0.2em}}
    \spyimageCAVE{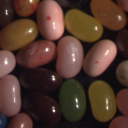} &
    \spyimageCAVE{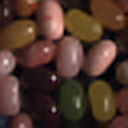}  &
    \spyimageCAVE{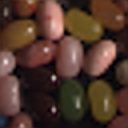}  &
    \spyimageCAVE{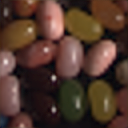}  \\

    GT & AWAN-SSPSR & SPSSR-AWAN & SSPSR-AWAN-Fusformer \\

    &
    \spyimageCAVE{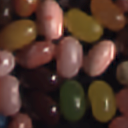}  &
    \spyimageCAVE{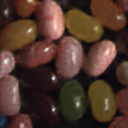}  &
    \spyimageCAVE{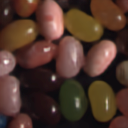}  \\

 & US3RN & SSFIN & Ours \\

\end{tabular}
    \caption{Visual comparison on CAVE test set for sampling of 2. For each method, the high-resolution result is displayed. Our method achieves sharper contours and correctly preserves color, while the other methods produce artifacts in the outlines of the jelly beans.}
    \label{fig:cave}
\end{figure*}

The CAVE dataset~\cite{yasuma2010generalized} consists of 32 scenes, where a variety of real-world materials and objects are captured. Each of these scenes is represented with HS data, covering wavelengths from 400 nm to 700 nm across 31 bands, where each band is stored as 16-bit grayscale PNG image. Additionally, an estimation of the spectral response is provided to generate the RGB representation of each scene. Among these scenes, 20 were used for the training set, 6 for the validation set, and the remaining 6 for the test set. Finally, each of the scenes has been split into patches of size 128x128 to reduce the memory cost during the training phase. 

Table~\ref{tab:results} shows the results on the CAVE dataset for 2, 4 and 8 sampling rates. The proposed method obtains the best results in all samplings and metrics. The second best position oscillates between SSFIN and US3RN, both specific methods for the purposed goal. The visual results displayed in Figure~\ref{fig:cave} are aligned with the conclusions of the quantitative comparison. Our method achieves sharper contours and correctly preserves color, while the other methods produce artifacts in the outlines of the jelly beans.

\subsection{Pavia Center}


\renewcommand{\spyimageCAVE}[2][0.22\linewidth]{%
  \setlength{\spyimagewidth}{#1}
  \setlength{\spyimageheight}{\spyimagewidth}
  \begin{tikzpicture}[spy using outlines={rectangle, magnification=2, size=0.5\spyimagewidth, connect spies,color=white}]
    \node[inner sep=0pt, anchor=south west] (image) at (0,0) {\includegraphics[width=\spyimagewidth]{#2}};
    \coordinate (spy point) at (0.35\spyimagewidth , 0.15\spyimagewidth );
    \coordinate (spy node) at (0.25\spyimagewidth, 0.75\spyimageheight);
    \spy on (spy point) in node at (spy node);
  \end{tikzpicture}%
}

\begin{figure*}[t]
    \centering
\begin{tabular}{c@{\hskip 0.2em} c@{\hskip 0.2em} c@{\hskip 0.2em} c@{\hskip 0.2em}}
    \spyimageCAVE{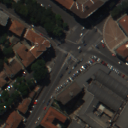} &
    \spyimageCAVE{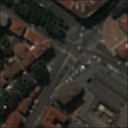}  &
    \spyimageCAVE{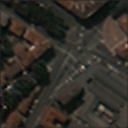}  &
    \spyimageCAVE{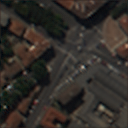}  \\

    GT & AWAN-SSPSR & SPSSR-AWAN & SSPSR-AWAN-Fusformer \\

    &
    \spyimageCAVE{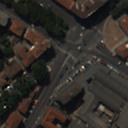}  &
    \spyimageCAVE{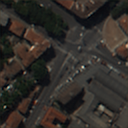}  &
    \spyimageCAVE{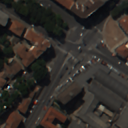}  \\

 & US3RN & SSFIN & Ours \\

\end{tabular}
    \caption{Visual comparison on Pavia test set for sampling of 2. For each method, the high-resolution result is displayed. Our method produces the most vivid colors as it is evidenced in the red car of the highlighted area. At the same time, we correctly preserve the image geometry, while the other approaches yield smoother results.}
    \label{fig:pavia}
\end{figure*}

Pavia Center is an image captured by ROSIS sensors during a flight campaign over Pavia, northern Italy. The data cover the spectral range from 430 nm to 860 nm, represented in 102 spectral bands. An estimation of the MS data (RGB and near infrared) for the same scene can be computed from the image. This image has been divided into crops of size 128x128, resulting in 40 examples. These resulting crops have been split into 32 for training, 4 for validation, and 4 for test.

As reported in Table~\ref{tab:results}, the proposed method outperforms the state of the art also in this dataset. We consistently achieve the best results on all metrics and sampling factors. Additionally, the visual results in Figure~\ref{fig:pavia} show that our model produces the most vivid colors—clearly seen in the red car within the highlighted area—and the sharpest edges. While US3RN and SSFIN correctly recover the geometry, non-specific SSSR methods tend to yield an overly smooth result, as demonstrated by the edges of the building and the street.

\subsection{NTIRE2020}


\newcommand{\spyeimagentire}[2][0.22\linewidth]{%
  \setlength{\spyimagewidth}{#1}
  \setlength{\spyimageheight}{\spyimagewidth}
  \begin{tikzpicture}[spy using outlines={rectangle, magnification=3, size=0.5\spyimagewidth, connect spies,color=white}]
    \node[inner sep=0pt, anchor=south west] (image) at (0,0) {\includegraphics[width=\spyimagewidth,trim={0 1.1cm 0 0},clip]{#2}};
    \coordinate (spy point) at (0.32\spyimagewidth , 0.795\spyimagewidth );
    \coordinate (spy node) at (0.25\spyimagewidth, 0.25\spyimageheight);
    \spy on (spy point) in node at (spy node);
  \end{tikzpicture}%
}

\begin{figure*}[t]
    \centering
\begin{tabular}{c@{\hskip 0.2em} c@{\hskip 0.2em} c@{\hskip 0.2em} c@{\hskip 0.2em}}
    \spyeimagentire{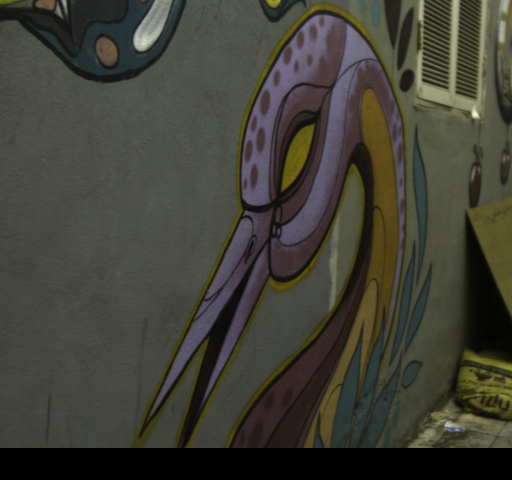} &
    \spyeimagentire{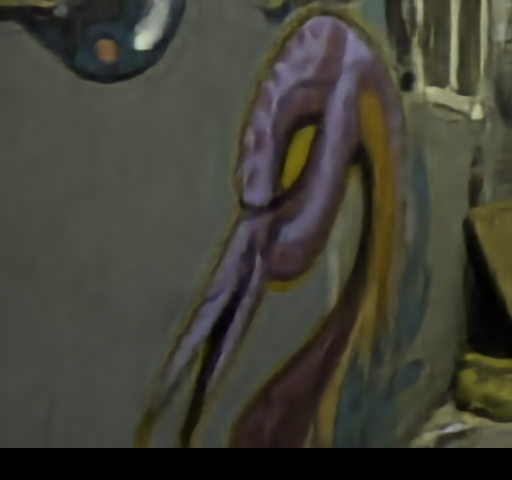}  &
    \spyeimagentire{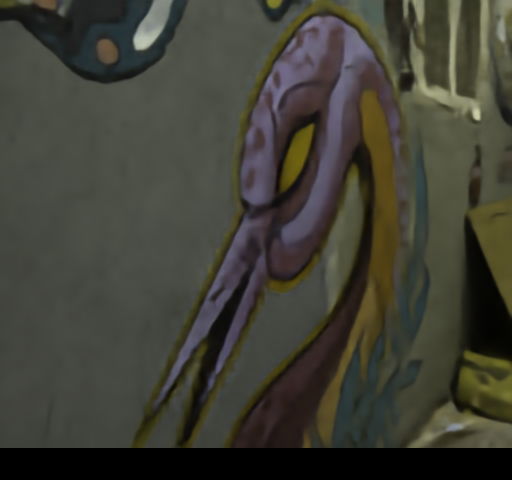}  &
    \spyeimagentire{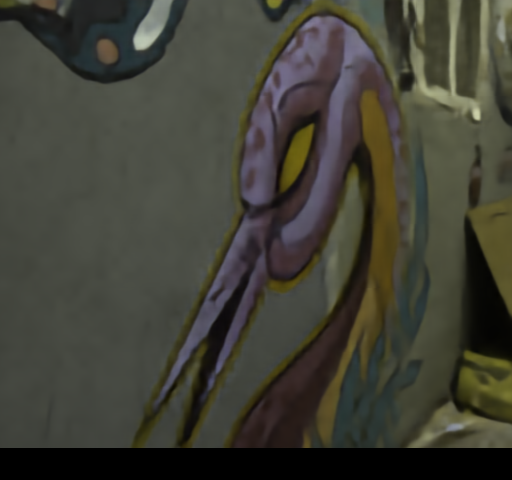}  \\

    GT & AWAN-SSPSR & SPSSR-AWAN & SSPSR-AWAN-Fusformer \\

    &
    \spyeimagentire{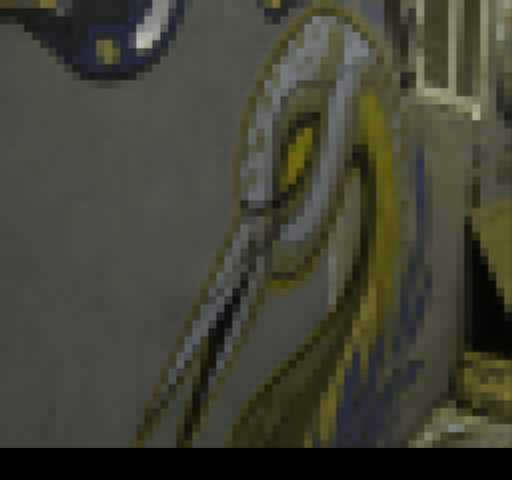}  &
    \spyeimagentire{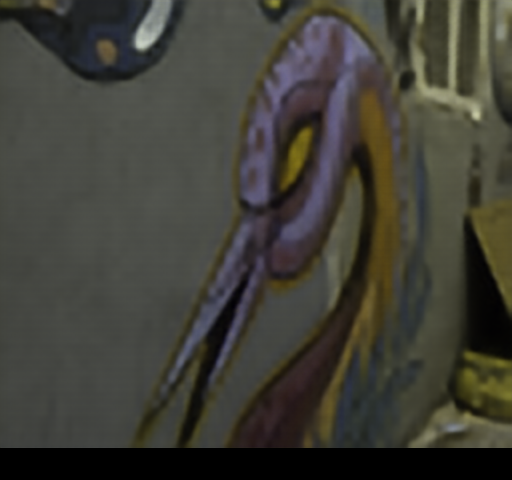}  &
    \spyeimagentire{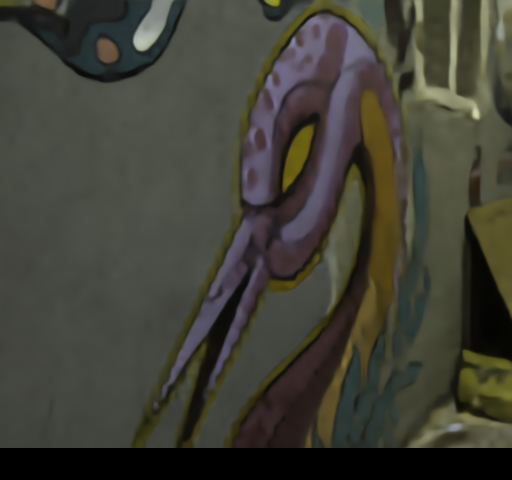}  \\

 & US3RN & SSFIN & Ours \\

\end{tabular}
    \caption{Visual comparison on NTIRE2020 test set for sampling of 8. For each method, the high-resolution result is displayed. As it can be observed in the highlighted area, US3RN and SSFIN are unable to recover chromaticity and fail to reconstruct the geometry. Non-specific SSSR methods offer a better spectral retrieval, but provide smooth solutions. In contrast, our method achieves superior geometry reconstruction and color preservation.}
    \label{fig:ntire2020}
\end{figure*}

The NTIRE2020 dataset was created for the NTIRE2020 challenge~\cite{arad2020ntire} on spectral reconstruction from RGB images. This challenge comprised clean and real-world tracks. In these experiments, we use the clean data, which include 450 training scenes and 10 examples for validation. We additionally divided the latter into validation and testing sets, with 5 images each. The data cover the wavelengths of 400-700 nm across 31 spectral bands. Also, an estimation of the spectral response is provided to obtain the RGB image. Each image has been split into crops of size 128x128 to reduce the memory cost during the training phase. 

As shown in Table~\ref{tab:results}, our method achieves the best metrics in all cases except for the sampling rate of 4, where it ranks second in ERGAS. 
The quantitative results are supported by the qualitative ones, illustrated in Figure~\ref{fig:ntire2020}. The figure displays the results for the sampling of 8 for one image of the test set. 
Although this sampling is challenging, our method achieves superior geometry reconstruction and color preservation, as evidenced in the highlighted area. In contrast, US3RN and SSFIN are unable to recover chromaticity and fail to reconstruct the geometry. Although non-specific SSSR methods can retrieve spectral information, they provide an oversmooth result.

\section{Ablation Study}\label{sec:ablation}

In this section, we evaluate the performance of the different operators designed for each component of the proposed approach. Specifically, we assess different architectures for the upsampling operator in the SR module, as well as for the downsampling and upsampling operators in the SSR module. Additionally, we explore the optimal values for several hyperparameters. These include the number of stages, the number of features, and the trade-off parameters in the training loss function. Finally, the impact of the proposed post-processing is studied. All these experiments have been carried out using the CAVE dataset with the data corresponding to a spatial sampling factor of 4.

\subsection{Spatial Upsampling}\label{sec:ablation-sr}

We consider four candidates to replace the spatial upsampling. Specifically, we contemplate two architectures that emulate the definition of the classic upsampling operators, as well as two others inspired by the back-projection algorithm. 

The upsampling operator of sampling $s$ is used to be modeled by a composition of an $s$-decimation and a low-pass filter. 
We have chosen a convolution to replace the low-pass filter and a transposed convolution for the $s$-decimation. The difference between the first two options is the application of these operations in a single step, or progressively in several steps according to the prime decomposition of the sampling factor $s$.

The two back-projection alternatives differ from each other in the chosen upsampling operators. As explained, these can be computed in a single step or in a progressive fashion.

We have compared these options with only the SR component of our proposed model. That is, we have a LR-MSI and a HR-MSI, as input and output data, respectively. As training loss, we only use the terms of the proposed loss function~\eqref{eq:loss} that involve the HR-MSI reference. In this case, the loss hyperparameters remain constant, specifically $\alpha_{SR}=0.3$. 

Table~\ref{tab:ablation_sr} displays the results across all metrics on the testing set. The classic-inspired option is denoted as {\it forward}, while the back-projection one as {\it BP-based}. It shows that the operator motivated by the back-projection algorithm combined with the progressive upsampling steps outperforms the other three options.
\begin{table}
\centering
\caption{Performance of different spatial upsampling methods for SR. The best metrics are highlighted in bold.}\label{tab:ablation_sr}
\begin{tabular}{l l | c c c c}
 & & PSNR$\uparrow$ & SSIM$\uparrow$ & SAM$\downarrow$ & ERGAS$\downarrow$ \\
 \hline
\multirow{2}{*}{Forward} & single step & 43.85 & 0.9729& 1.75 & 5.23\\
 & progressive & 43.88 & 0.9732& 1.77&5.21  \\\hline
\multirow{2}{*}{BP-based} & single step & 43.89 & 0.9731& 1.77&5.22 \\
 & progressive & \textbf{43.97} & \textbf{0.9750}& \textbf{1.67} &\textbf{5.18}\\
\end{tabular}
\end{table}

\subsection{Spectral Upsampling and Downsampling}

We explored three approaches to replace the spectral upsampling and downsampling processes. All of these approaches employ an MLP to estimate the transition between HS and MS representations, or vice versa.

The first approach estimates the same MLP weights for all pixels. In contrast, the other two approaches divide the image into $M$ clusters, with different weights learned for each cluster. This means that an adapted MLP is applied depending on the cluster to which a pixel belongs. Two methods for cluster estimation were explored. On the one hand, the $k$-means algorithm was employed, with centroids computed using 25$\%$ of the training data selected randomly. On the other hand, a learnable architecture was used to determine the clusters.

For these comparisons, we only applied the SSR unit of our model. Therefore, we have a LR-MSI as input and a LR-HSI as output. The loss function only considers the terms in \eqref{eq:loss} that depend on the LR-HSI, with a constant $\alpha_{SSR}=0.2$. Table~\ref{tab:ablation_ssr} presents the metrics on the test set, showing that the approach using learned clusters consistently achieves the best results on all metrics.

\begin{table}
\centering
\caption{Comparison of different spectral upsampling and downsampling approaches for SSR. Best metrics are highlighted in bold.}\label{tab:ablation_ssr}
\begin{tabular}{r |c c c c}
 & PSNR$\uparrow$ & SSIM$\uparrow$ & SAM$\downarrow$ & ERGAS$\downarrow$ \\
 \hline
No clusters & 39.25 & 0.9789& 7.55 & 11.08 \\
K-means clusters & 39.01 & 0.9802& 6.87 & 14.22 \\
Learned clusters & \textbf{39.40} & \textbf{0.9821}& \textbf{6.64} & \textbf{7.80} \\
\end{tabular}
\end{table}
\subsection{Effect of the number of stages and features}

\begin{table}
\centering
\caption{Results of the model trained with different number of stages.The best results are highlighted in bold.}\label{tab:ablation_stages}
\begin{tabular}{c | c c c c}
Stages & PSNR$\uparrow$ & SSIM$\uparrow$ & SAM$\downarrow$ & ERGAS$\downarrow$ \\
\hline
2 & 37.10 & 0.9400 & 9.18 & 8.64 \\
4 & \textbf{37.11} & \textbf{0.9401} & \textbf{9.17} & \textbf{8.28} \\
6 & 37.04 & 0.9395 & 9.23 & 8.66 \\
8 & 36.76 & 0.9400 & 9.32 & 8.58 \\
\end{tabular}
\end{table}

\begin{table}
\centering
\caption{Results of the model trained with different number of features.The best results are highlighted in bold.}\label{tab:ablation_features}
\begin{tabular}{c | c c c c}
Features & PSNR$\uparrow$ & SSIM$\uparrow$ & SAM$\downarrow$ & ERGAS$\downarrow$ \\
\hline
16 & 36.83 & 0.9323 & 9.36 & 8.53 \\
32 & 36.92 & 0.9398 & 9.28 & 8.71 \\
64 & 37.11 & 0.9401 & 9.17 & \textbf{8.28} \\
128 & \textbf{37.21} & \textbf{0.9417} & \textbf{9.15} & 8.45 \\
256 & 36.64 & 0.9401 & 9.96 & 9.03 \\
\end{tabular}
\end{table}

The number of stages $K$ and the number of deep features used in each module have a significant impact on the performance of our approach. By this number of features, we refer to the number of channels used in the sequence of residual blocks of the learning-based proximity operators. We chose these two parameters to be the same in the three SR, SSR and image fusion iterative processes. 
In this section, we show the effect of setting different values of these two parameters. 
For both purposes, we evaluate the model without the post-processing step. Also, the loss hyperparameters remain constant during the entire training phase. Specifically, $\alpha_{SR}=0.5$, $\alpha_{SSR}=0.2$ and $\alpha_{Fus}=0.3$.

Regarding the number of stages, we trained our model with 2, 4, 6, and 8 stages. In all cases, we fixed the number of features to 64. 
The results are shown in Table \ref{tab:ablation_stages}, where we can observe that the best metrics are achieved when we use 4 stages. 

On the other hand, we fixed the number of stages and determined the optimal number of features. In particular, we tested our approach with 16, 32, 64, 128 and 256 deep features, setting $K=4$. Table \ref{tab:ablation_features} shows that the best results are obtained when we establish 128 features.

\subsection{Loss Hyperparemeters}

\begin{table}\label{tab:scheduled_loss}
\centering
\caption{Comparison of the combination of different hyperparemters in the training loss. The best metrics are highlighted in bold.}
\begin{tabular}{c | c c c c}
 & PSNR$\uparrow$ & SSIM$\uparrow$ & SAM$\downarrow$ & ERGAS$\downarrow$ \\
 \hline
$\text{comb}_{1,1,1} $& 37.11 & 0.9418& 9.30 & 8.51 \\
$\text{comb}_{1,1,2}$ & 36.89 & 0.9414& 9.49 & 8.53 \\
$\text{comb}_{1,2,1}$ & 36.96 & 0.9411& 9.57 & 8.50 \\
$\text{comb}_{2,1,1}$ & \textbf{37.25} & \textbf{0.9429}& \textbf{9.11} & 8.27 \\
$\text{comb}_{2,1,2}$ & 36.99 & \textbf{0.9429}& 9.39 & 8.39 \\
$\text{comb}_{2,2,1}$ & 36.93 & 0.9416& 9.37 & \textbf{8.26} \\
$\text{comb}_{2,2,2}$ & 37.04 & 0.9407& 9.52 & 8.64 \\
\end{tabular}
\end{table}

The proposed loss \eqref{eq:loss} involves three hyperparameters that are updated during the first training phase. In this phase, we optimize our networks during 1000 epochs, and these hyperparameters are updated at epochs 300 and 600. To find a suitable combination of these values, we have tested several possibilities. For each hyperparameter, two sets of values have been evaluated. Specifically, for $\alpha_{SR}$ we consider $\{1,1,0.5\}$ and $\{2,0.5,0\}$; for $\alpha_{SSR}$ we consider $\{1,1,0.5\}$ and $\{2,0.5,0\}$; and for $\alpha_{Fusion}$ we consider $\{0.5,1,1\}$ and $\{0,0.5,2\}$. Note that in each set, the first element represents the initial value, the second one the updated value at epoch 300, and the third one refers to the value we set from epoch 600. In total, we have 8 possible combinations. Each of these has been trained on the CAVE dataset with a sampling factor of 4. Also, for this experiment we do not apply the post-processing module. 

We denote each combination as $\text{comb}_{i,j,k}$, where $i,j,k\in\{1,2\}$, correspond to the first or second option for  $\alpha_{SR}$, $\alpha_{SSR}$, and $\alpha_{Fus}$, respectively. 
Table~\ref{tab:scheduled_loss} shows the metrics for each option on the CAVE test set. We achieve the best performance with $\text{comb}_{2,1,1}$. This corresponds to $\alpha_{SR}=\{2,0.5,0\}$, $\alpha_{SSR}=\{1,1,0.5\}$, and $\alpha_{Fus}=\{0.5,1,1\}$. Therefore, in the initial epochs it works best to give more weight to the loss terms associated with the SR and the SSR outcomes, and progressively decrease them as the training advances. On the contrary, it is preferred to escalate with the training epochs the weight involving the fused image.

\subsection{Impact of post-processing}

\begin{table}
\centering
\caption{Quantitative evaluation with and without post-processing. The best results are highlighted in bold.}\label{tab:ablation_post}
\begin{tabular}{r | c c c c }
 & PSNR$\uparrow$ & SSIM$\uparrow$ & SAM$\downarrow$ & ERGAS$\downarrow$ \\
 \hline
Without post-processing & 37.25 & 0.9395& 9.11 & \textbf{8.27} \\
With post-processing & \textbf{37.40} & \textbf{0.9427}& \textbf{8.96} & 8.28 \\
\end{tabular}
\end{table}

Finally, to conclude the study of the model, we examine the effect of the post-processing module. This post-processing allows us to further enhance the geometry and chromaticity of the fused result. We have compared the performance of the post-processing quantitatively. Table~\ref{tab:ablation_post} displays the metrics on the test set with and without this component. A significant improvement is achieved in all metrics except for ERGAS, where the loss is negligible. 

\section{Conclusions}\label{sec:conclusions}

In this work, we have addressed the spatio-spectral super-resolution problem by decomposing it into spatial super-resolution, spectral super-resolution, and image fusion subtasks. The SR and SSR processes operate in parallel on the LR-MSI, and their outputs are subsequently fused to produce the final HR-HSI.

Each subtask has been approached from an unfolding perspective, leading to a flexible and interpretable architecture. The design of all involved operators has been carefully tailored to the specific requirements of each subtask, and their effectiveness has been assesses through ablation studies.

We have further incorporated a post-processing module that improves geometric and spectral consistency. This stage leverages image self-similarity by means of multi-head attention layers that operate at pixel level. The attention weights are restricted to the top 10$\%$ of the most similar pixels with a given neighborhood. This allows us to efficiently capture long-range dependencies in the high-resolution domain..

The performance of the proposed end-to-end model has been compared with state-of-the-art approaches for SSSR. We have also included the results obtained by combining models designed separately for SR, SSR, and image fusion. The evaluation has been conducted on different datasets and sampling factors. In all cases, both quantitative and qualitative results exhibits the superior performance of our proposal. 

For future work, we plan to investigate how nonlocal operations can be more effectively integrated into the proximity operators, while maintaining computational efficiency and accuracy. This could potentially eliminate the need for a separate post-processing step. In addition, we aim to explore more efficient strategies for computing similarities across the entire image.

\begin{bibliography}{biblio}
\bibliographystyle{IEEEtran}
\end{bibliography}

\end{document}